\documentclass[twocolumn]{aastex631}

\usepackage{amsmath}
\usepackage{color}

\begin{document}

\title{Baryon-induced collapse of dark matter cores into supermassive black holes}

\author[0000-0002-5862-8840]{C. R. Argüelles}
\affiliation{Instituto de Astrof\'isica de La Plata, UNLP-CONICET, Paseo del Bosque s/n B1900FWA La Plata, Argentina}
\affiliation{ICRANet, Piazza della Repubblica 10, I-65122 Pescara, Italy}

\author[0000-0003-4904-0014]{J. A. Rueda}
\affiliation{ICRANet, Piazza della Repubblica 10, I-65122 Pescara, Italy}
\affiliation{ICRANet-Ferrara, Dip. di Fisica e Scienze della Terra, Universit\`a degli Studi di Ferrara, Via Saragat 1, I--44122 Ferrara, Italy}
\affiliation{ICRA, Dipartamento di Fisica, Sapienza Universit\`a  di Roma, Piazzale Aldo Moro 5, I-00185 Rome, Italy}
\affiliation{Department of Physics and Earth Science, 
University of Ferrara, Via Saragat 1, I-44122 Ferrara, Italy}
\affiliation{INAF, Istituto di Astrofisica e Planetologia Spaziali, Via Fosso del Cavaliere 100, 00133 Rome, Italy}

\author[0000-0003-0829-8318]{R.~Ruffini}
\affiliation{ICRANet, Piazza della Repubblica 10, I-65122 Pescara, Italy}
\affiliation{ICRA, Dipartamento di Fisica, Sapienza Universit\`a  di Roma, Piazzale Aldo Moro 5, I-00185 Rome, Italy}
\affiliation{Universit\'e de Nice Sophia-Antipolis, Grand Ch\^ateau Parc Valrose, Nice, CEDEX 2, France}
\affiliation{INAF, Viale del Parco Mellini 84, 00136 Rome, Italy}

\email{carguelles@fcaglp.unlp.edu.ar; jorge.rueda@icra.it; ruffini@icra.it}

\date{Received date /Accepted date }

\begin{abstract}
    Non-linear structure formation for fermionic dark matter particles leads to dark matter density profiles with a degenerate compact core surrounded by a diluted halo. {For a given fermion mass, the core has a critical mass that collapses into a supermassive black hole (SMBH). Galactic dynamics constraints suggest a $\sim 100$ keV/$c^2$ fermion, which leads to $\sim 10^7 M_\odot$ critical core mass. Here, we show that baryonic (ordinary) matter accretion drives an initially stable dark matter core to SMBH formation and determine the accreted mass threshold that induces it. Baryonic gas density $\rho_b$ and velocity $v_b$ inferred from cosmological hydro-simulations and observations produce sub-Eddington accretion rates triggering the baryon-induced collapse in less than a Gyr. This process produces active galactic nuclei in galaxy mergers and the high-redshift Universe. For TXS 2116--077, merging with a nearby galaxy, the observed $3\times 10^7 M_\odot$ SMBH, for $Q_b = \rho_b/v_b^3 = 0.125 M_\odot/(100\,\text{km/s pc})^3$, forms in $\approx 0.6$ Gyr, consistent with the $0.5$--$2$ Gyr merger timescale and younger jet.  For the farthest central SMBH detected by the \textit{Chandra} X-ray satellite in the $z= 10.3$ UHZ1 galaxy observed by the James Webb Space Telescope (\textit{JWST}), the mechanism leads to a $4\times 10^7 M_\odot$ SMBH in $87$--$187$ Myr, starting the accretion at $z=12$--$15$. The baryon-induced collapse can also explain the $\approx 10^7$--$10^8 M_\odot$ SMBHs revealed by the JWST at $z\approx 4$--$6$. After its formation, the SMBH can grow to a few $10^9 M_\odot$ in timescales shorter than a Gyr via sub-Eddington baryonic mass accretion.}
\end{abstract}



\section{Introduction}\label{sec:1}

The standing problem of early formation and growth of SMBHs poses a significant challenge to our current cosmological understanding. Directly related open issues include which is the main channel for the formation of the SMBH seeds in the high-$z$ Universe and how such a BH-seed mass correlates with the hosting halo mass in the given cosmological evolution  \citep[see, e.g.,][and references therein]{2020ARA&A..58...27I,2021NatRP...3..732V,2023arXiv230702531A,2023ApJ...953L..29L,2023arXiv230515458B,2023arXiv230812331P}. Standard baryonic channels are based on the growth of BH seeds of $\lesssim 10^2 M_\odot$ from the collapse of population III stars \citep{2001ApJ...551L..27M,2016ApJ...824..119H} or of $10^4$--$10^5 M_\odot$ from the direct collapse of gaseous configurations in a plethora of physical setups \citep{2006MNRAS.370..289B,2008MNRAS.387.1649B, 2017ApJ...842L...6W,2022MNRAS.514.5583Z,2022Natur.607...48L}. These models are in tension with observations since the BH seeds of $10^2$--$10^5 M_\odot$ cannot grow to $\sim 10^9 M_\odot$ sufficiently fast to explain observations at high redshift, e.g., $z \sim 6$ unless ad-hoc or extreme assumptions {are made} (e.g., sustained Eddington or super-Eddington accretion rates for long cosmological times \citealp{2021ApJ...923..262Y}) {limiting} the generality of hydrodynamic numerical simulations \citep[see, e.g.,][]{2022MNRAS.514.5583Z}. Such tension is increasing with the daily \textit{JWST} observations and new data of quasars at high redshift (e.g., $z\sim 6$--$7$ \citealp{2021ApJ...923..262Y,2023arXiv230904614Y}), exploring the faint end of the {SMBH} luminosity function, unveiling a larger population than previously thought (see  \citealp{2022A&A...666A..17G,2023arXiv230801230M,2023ApJ...954L...4K} and \citealp{2023ARA&A..61..373F} for a recent review).

Recently, a new SMBH formation channel has been proposed based on the gravitational collapse of dark matter cores of $\sim 10^7 M_\odot$, made of $\sim 100$ keV/$c^2$ fermions \citep{2023MNRAS.523.2209A}. The cores are those of the \textit{dense core}-\textit{diluted halo} dark matter galactic profiles predicted by the Ruffini-Arg\"uelles-Rueda (RAR) model \citep{2015MNRAS.451..622R,2018PDU....21...82A} (see next section for details). These core-halo density profiles explain a variety of galactic observables, including the flat rotation curves \citep{2018PDU....21...82A,2023ApJ...945....1K}, galactic universal relations \citep{2019PDU....24..278A,2023ApJ...945....1K}, and the motion of the innermost stars near the Milky Way's center \citep{2020A&A...641A..34B,2021MNRAS.505L..64B,2022MNRAS.511L..35A}. Paper \citep{2023MNRAS.523.2209A} showed that starting from the BH seeds produced by the collapse of these dark matter cores, SMBHs of a few $10^9 M_\odot$ can form in less than a Gyr by accreting baryonic matter at sub-Eddington rates.

The appeal of the above dark matter core collapse scenario to solve the problem of formation and growth of SMBHs at high $z$ has led us to inquire further about it. Thus, we answer in this Letter three highly relevant questions left open in \citet{2023MNRAS.523.2209A}: 1) how does the dark matter core of fermions reach the point of gravitational collapse?; 2) how long does it take to the core to reach that point? and 3) are those conditions attainable in astrophysical and cosmological setups?

The answer to these questions arises from the fact that galaxy halos are not only made of dark matter but also have baryonic, i.e., ordinary matter. The baryonic matter infall into the {potential well of the dark matter core} modifies its equilibrium state. We calculate those new equilibrium configurations with the presence of baryonic matter and establish the existence of a critical mass for the gravitational collapse that depends upon the amount of baryonic matter settled in the core. The gravitational instability occurs because baryons contribute to the system energy density but not to the pressure, which is governed by the fermion degeneracy pressure in the dense dark matter core. We determine the threshold amount of baryonic mass and the minimum mass of the initial dark matter core leading to the SMBH formation {(section \ref{sec:2})}. We then show that SMBHs of $10^7 M_\odot$ can be formed from the baryon-induced collapse of dark matter cores in timescales shorter than a Gyr, for baryonic inner densities and velocities as obtained in cosmological simulations and observations {(section \ref{sec:3}). Section \ref{sec:4} examines specific applications in galaxy mergers and the high-$z$ Universe. By examining the data from the \textit{Chandra} satellite and William Herschel Telescope of the Seyfert galaxy TXS 2116--077, we show that the merging timescale and jet lifetime agree with the SMBH originating from the baryon-induced collapse and that the inferred baryonic environment conditions for its occurrence agree with those observed. The baryon-induced collapse can also explain the formation of the $\sim 10^7$-$10^8 M_\odot$ farthest quasar, observed by the \textit{Chandra} satellite at the center of the \textit{JWST}-detected galaxy UHZ1 at $z=10.3$. The same conclusions apply to the \textit{little red dots}, the SMBHs at $z\approx 4$--$6$ also observed by \textit{JWST}}.

\section{The baryon-induced collapse}\label{sec:2}

\begin{figure}
    \centering
    \includegraphics[width=\hsize,clip]{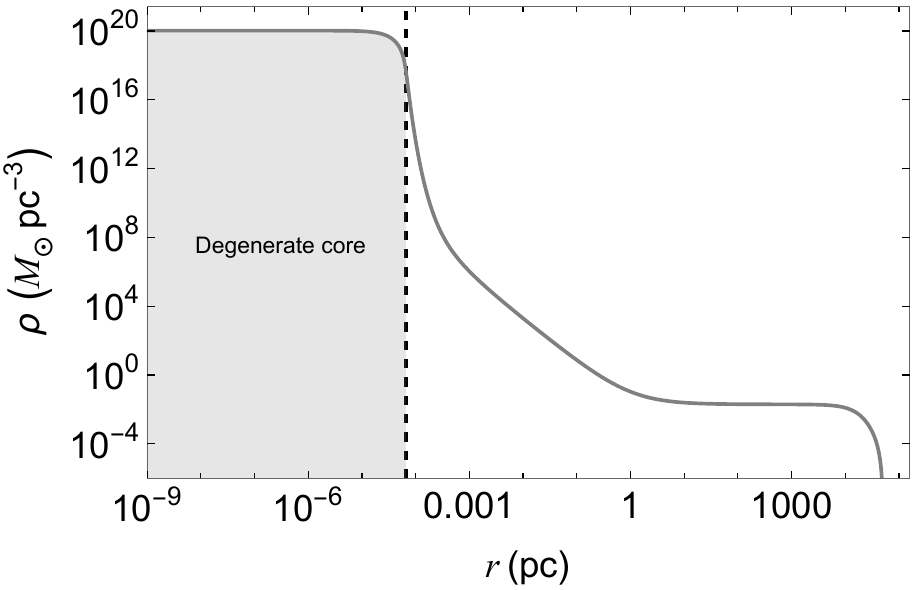}
    \caption{Pure-dark-matter equilibrium configuration with central density $\rho = 0.0068$ g cm$^{-3}=1.01\times 10^{20} M_\odot$ pc$^{-3}$, for a fermion $m = 100$ keV/$c^2$. The total halo mass is $5\times 10^{11} M_\odot$. The degenerate quantum core (filled gray region) has a mass $M_{\rm dm} = 2.03 \times 10^7 M_\odot$ and radius $R_c = 6.67\times 10^{-5}$ pc.}
    \label{fig:profile}
\end{figure}

The RAR model treats the dark matter in galaxies as a self-gravitating system of fermions at finite temperatures in equilibrium, so the general relativity field equations set the structure. The dark matter equation of state obeys Fermi-Dirac statistics and includes a particle energy cutoff, which determines the galaxy's finite size (see \citealp{2018PDU....21...82A}; also \citealp{2015PhRvD..92l3527C} for a Newtonian approach). The formation and stability of core-halo profiles with a Fermi-Dirac-like distribution function are predicted in structure formation scenarios of maximum entropy production principle \citep{1998MNRAS.300..981C,2021MNRAS.502.4227A,2023MNRAS.523.2209A}. The equilibrium configuration is characterized by the segregation of the fermion physical regimes along the galaxy: a quantum degenerate core of nearly uniform density, followed by an intermediate semi-classical regime where the density falls off abruptly, followed by a plateau. Then, it follows a Boltzmann-like regime in the outer halo where the density falls off as a power-law, followed by an exponential decrease determining the galaxy border. Figure \ref{fig:profile} shows an example of a galactic dark matter profile for $m = 100$ keV/$c^2$.

We focus hereafter on the equilibrium state of the core where fermions are in a degenerate regime. We can analyze the stability of this core by solving the Einstein equations for the simpler equation of state of a fermion gas at zero temperature. The energy density and pressure of such pure-dark-matter fermion gas are, respectively, ${\cal E} = {\cal E}_{\rm dm}$ and $P = P_{\rm dm}$, where
\begin{subequations}\label{eq:EOS}
    \begin{align} 
    {\cal E}_{\rm dm} &= \frac{m c^2}{8 \pi^2 \lambda^3}[\sqrt{1+x^2}(x + 2 x^3) - {\rm arcsinh}(x)],\\
    P_{\rm dm} &=\frac{m c^2}{24 \pi^2 \lambda^3}[x \sqrt{1+x^2}(2 x^2-3) + \frac{3}{8}{\rm arcsinh}(x)],
\end{align}
\end{subequations}
being $\lambda = \hbar/(m c)$ the particle Compton wavelength and $x = p_F/(m c)$ the dimensionless Fermi momentum, which sets the particle rest-mass density, $\rho_{\rm dm} = m\,n_{\rm dm} = m\,x^3/(3 \pi^2 \lambda^3)$. The equilibrium configurations satisfy the Einstein equations in spherical symmetry for a perfect fluid, which can be written in the Tolman-Oppenheimer-Volkoff form \citep{1939PhRv...55..374O}
%
    \begin{equation}\label{eq:TOV}
    \frac{dM}{dr} = 4 \pi r^2 \rho ,\quad
    \frac{dP}{dr} = - G\frac{(\rho + P/c^2)(4 \pi r^3 P/c^2 + M)}{r (r - 2 G M/c^2)}.
\end{equation}
%
where $M$ is the gravitational mass measured by an observer at rest at infinity. For instance, the solution of Eqs. (\ref{eq:TOV}) for a pure-dark-matter equilibrium configuration with central density $\rho = 6.8\times 10^{-3}$ g cm$^{-3}\approx 10^{20} M_\odot$ pc$^{-3}$ leads to the density profile highlighted by the filled gray region in Fig. \ref{fig:profile}, with mass $M_{\rm dm} = 2.03 \times 10^7 M_\odot$ and radius $R_c = 6.67\times 10^{-5}$ pc.

Along a sequence of equilibrium configurations with increasing central density, the turning point, i.e., where $\partial M/\partial \rho_c = 0$, being $\rho_c = {\cal E}_c/c^2$, and the subscript $c$ stands for values at the center ($r=0$), sets the configuration of critical density and corresponding critical mass over which the core becomes unstable against gravitational collapse
\citep{1939PhRv...55..374O}. The critical mass of pure-dark-matter cores is \citep{2021MNRAS.502.4227A,2023MNRAS.523.2209A} 
\begin{equation}\label{eq:Mcrit}
M^{(0)}_{\rm crit}\approx 0.38 \frac{m_{\rm Pl}^3}{m^2} = 6.27 \times 10^7 \left( \frac{\text{100 keV}/c^2}{m}  \right)^2 M_{\odot},
\end{equation}
where $m_{\rm Pl} = \sqrt{\hbar c/G} = 2.18 \times 10^{-5}$ g is the Planck's mass. For $m = 100$ keV/$c^2$ fermions, Eq. (\ref{eq:Mcrit}) tells that the gravitational collapse of the dark matter core would lead to a BH mass $M_{\rm BH} = M^{(0)}_{\rm crit}= 6.27\times 10^7 M_\odot$.

\begin{figure}
    \centering
    \includegraphics[width=\hsize,clip]{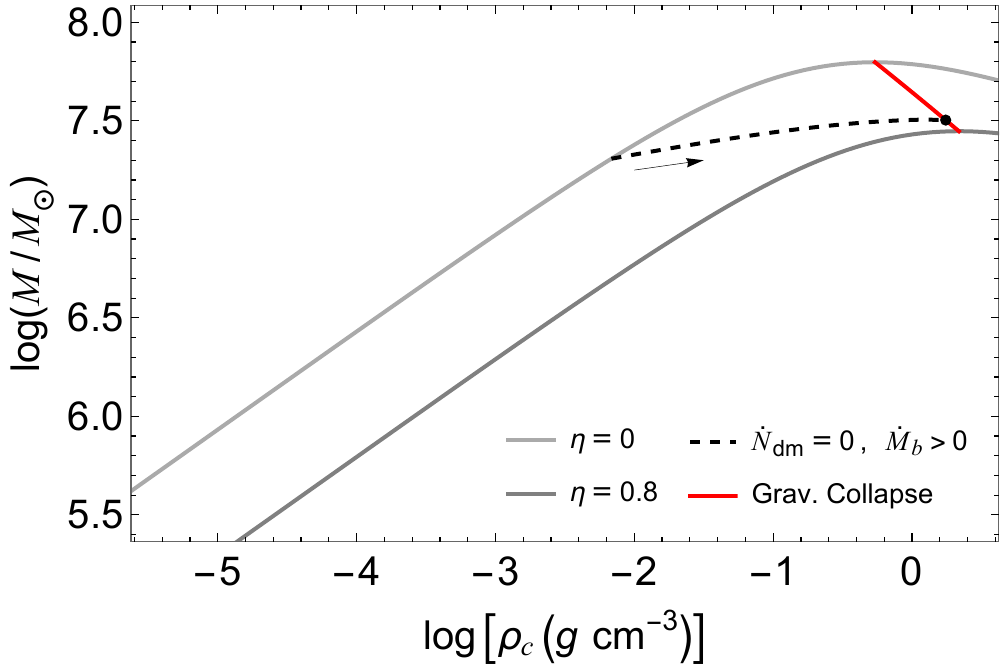}
    \caption{Dark matter+baryonic equilibrium configurations for $m = 100$ keV/$c^2$. The black dashed curve shows the equilibrium configurations at a constant total fermion number. The lighter and darker gray curves are the equilibrium sequences $\eta = 0$ and $0.8$. The red curve is the secular instability limit for gravitational collapse. The black-dashed sequence starts with a pure dark matter core ($\chi = 0$) of $M_{\rm dm} = 2.03 \times 10^7 M_\odot$ and radius $R_c = 6.67\times 10^{-5}$ pc, i.e., the degenerate core of the configuration shown in Fig. \ref{fig:profile}. As baryons sink in the core, the configuration follows the black-dashed sequence to the right (see the arrow) until it reaches the critical mass for gravitational collapse (black dot). The critical configuration has a baryon-to-dark-matter mass ratio $\chi_{\rm BH}\approx 0.56$, so baryons contribute $36\%$ and the dark matter $64\%$ to the critical mass, $M_{\rm BH} \approx 3.19\times 10^7 M_\odot$, i.e., $M_{\rm dm, crit} = 2.04\times 10^7 M_\odot$ and $M_{b, \rm crit} \approx 1.15 \times 10^7 M_\odot$.}
    \label{fig:mrhoplane}
\end{figure}

We are now interested in the modified equilibrium state of the core when it contains a composition of degenerate dark matter fermions and ordinary/baryonic matter. For this task, we model the baryon content as dust, so the equation of state is now given by ${\cal E} = {\cal E}_{\rm dm} + {\cal E}_b$, $P = P_{\rm dm}$, 
where ${\cal E}_b = c^2 \rho_b$ is the energy density of baryons, being $\rho_b$ the baryon rest-mass density. The system of equations (\ref{eq:TOV}) for this new equation of state can be integrated once we set a prescription to calculate $\rho_b$. We make the ansatz that the baryon rest-mass density follows the dark matter one, i.e., $\eta \equiv \rho_b/\rho_{\rm dm}={\rm constant}$. We construct the new equilibrium configuration sequences ($M$-$\rho_c$ sequences) for different values of $\eta$. In Fig. \ref{fig:mrhoplane}, we show the region of equilibrium configurations of dark matter+baryon cores for $0\leq \eta\leq 0.8$, in the case $m=100$ keV/$c^2$. For every given ratio, the sequence turning point gives a critical mass configuration where $\partial M/\partial \rho_c = 0$. This procedure leads to the sequence of critical mass configurations given by the red curve in Fig. \ref{fig:mrhoplane}. For the astrophysical analysis, using the baryon-to-dark-matter mass fraction $\chi = M_b/M_{\rm dm}$ is more convenient. The pure-dark-matter core case ($\chi=0$) is the first point (from top to bottom) of the critical configurations given by the red curve in the figure. The following analytic function fits the critical masses of the red curve:
\begin{equation}\label{eq:Mcritchi}
    M_{\rm BH} \equiv M_{\rm crit}\approx \frac{M^{(0)}_{\rm crit}}{1 + 1.466 \chi_{\rm BH} +0.458 \chi_{\rm BH}^2},
\end{equation}
where $\chi_{\rm BH} \equiv \chi_{\rm crit} = (M_b/M_{\rm dm})_{\rm crit} \approx M_{b,\rm crit}/M_{\rm dm}$, being $M_{b,\rm crit}$ the baryon mass at the point of critical mass, and $\dot{M}_{\rm dm} \approx 0$ at one-percent level along any constant-fermion-number track. The critical mass sets the SMBH mass formed from the collapse. It is worth emphasizing that Eq. (\ref{eq:Mcritchi}) is valid for any fermion mass $m$, for baryon-to-dark-matter mass ratios $0\leq \chi_{\rm BH}\lesssim 0.8$.

To exemplify the baryon-induced collapse, we show in Fig. \ref{fig:mrhoplane} a sequence of constant total dark matter particle number, $N_{\rm dm}$, while the baryon number increases along the direction the arrow indicates. The sequence starts with a pure-dark-matter core ($\chi = 0$) of gravitational mass $M_{\rm dm} = 2.03 \times 10^7 M_\odot$, corresponding to a dark matter particle number $N_{\rm dm} = 1.81\times 10^{67}$. The core is surrounded by a diluted halo of $M_h \sim 10^{11} M_\odot$ as predicted by the RAR model (see Fig. \ref{fig:profile}). According to Eq. (\ref{eq:Mcrit}), the initial dark matter core is stable, i.e., $M_{\rm dm} < M^{(0)}_{\rm crit}$. However, the sequence intersects the instability sequence (red curve) at a critical mass of $M_{\rm BH} = 3.19\times 10^7 M_\odot$, for a critical baryon-to-dark-matter mass ratio $\chi_{\rm BH}= 0.56$.

Therefore, the presence of baryons induces the collapse of an otherwise stable dark matter core, and the critical mass can be less than half of the critical mass of a pure-dark-matter core. But can dark matter cores gain those amounts of baryons in realistic astrophysical situations, at or about the moment of halo formation? 

\section{Baryon gravitational capture and accretion rate}\label{sec:3}

To answer the above question, we analyze the astrophysical situation where the dark matter core captures baryons from an inner halo environment in the high-$z$ universe. The core capture baryons of rest-mass density $\bar{\rho}_b$ and speed $v_b$ at a rate
\begin{equation}\label{eq:Mdotb}
    \dot{M}_b = \pi R_{\rm cap}^2 \bar{\rho}_b v_b,
\end{equation}
where $R_{\rm cap} = 2 G M/v_b^2$ is the gravitational capture radius. It is worth emphasizing that $\bar{\rho}_b$ is the density of baryons at $r=R_{\rm cap}$, which is expected to be much smaller than $\rho_b$ inside the dark matter core. We solve the evolution equation $\dot{M} = \dot{M}_{\rm dm} + \dot{M}_b$, which we can approximate to $\dot{M}\approx \dot{M}_b$ since we are assessing the evolution in a sequence $\dot{N}_{\rm dm} = 0$ and $\dot{M}_{\rm dm}\approx 0$ at the percent level. By integrating this equation, we obtain the mass and baryon-to-dark-matter mass ratio evolution
{
\begin{align}
    M (t)&= \frac{M_{\rm dm}}{1-\bar{t}},\quad \chi(t) = \frac{M_b(t)}{M_{\rm dm}} =  \frac{\bar{t}}{1-\bar{t}},\quad  \bar{t}\equiv\frac{t}{\tau}\label{eq:Mvst}\\
   \tau &\equiv \frac{1}{4\pi G^2 Q_b M_{\rm dm}} \approx \frac{421}{\bar{Q}_b}\left(\frac{10^7 M_\odot}{M_{\rm dm}}\right)\,\,\text{Myr},\label{eq:tau}
\end{align}
where $Q_b = \bar{\rho}_b/v_b^3$ and $\bar{Q}_b \equiv Q_b/[(M_\odot/{\rm pc}^3)/(100\,{\rm km/s})^3]$}, and we recall that $M_{\rm dm} = M(t=0)$ is the initial mass which is that of the pure-dark-matter core. Equation (\ref{eq:Mvst}) tells that the core reaches the critical mass for BH formation in a time
\begin{equation}\label{eq:tcrit}
    \bar{t}_{\rm BH} = \frac{\chi_{\rm BH}}{1+\chi_{\rm BH}} = 1 - \frac{M_{\rm dm}}{M_{\rm BH}}.
\end{equation}
By replacing Eqs. (\ref{eq:Mcritchi}) and (\ref{eq:Mvst}) into Eq. (\ref{eq:tcrit}), we obtain an algebraic equation for $\chi_{\rm BH}$ (or for $\bar{t}_{\rm BH}$) that we solve with the aid of a Padé approximant and obtain
\begin{equation}\label{eq:chiBH}
\chi_{\rm BH} \approx -0.83-\frac{0.09}{\mu} +\frac{\sqrt{0.008 + 0.896 \mu - 0.054\mu^2}}{\mu},
\end{equation}
where $\mu \equiv M_{\rm dm}/M^{(0)}_{\rm crit}$, accurate at the one-percent level.

\begin{figure*}
    \centering
    \includegraphics[width=0.48\hsize,clip]{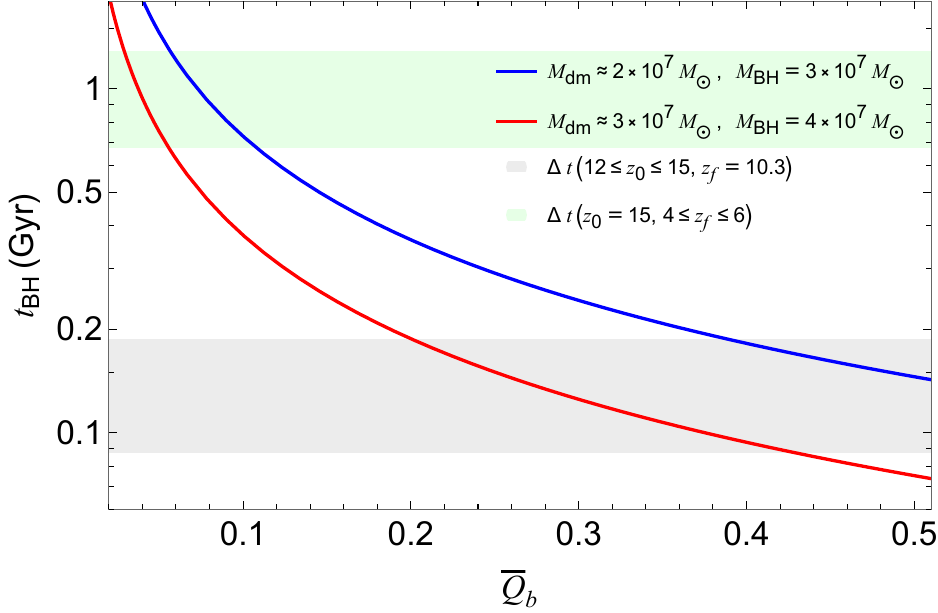}\,\includegraphics[width=0.48\hsize,clip]{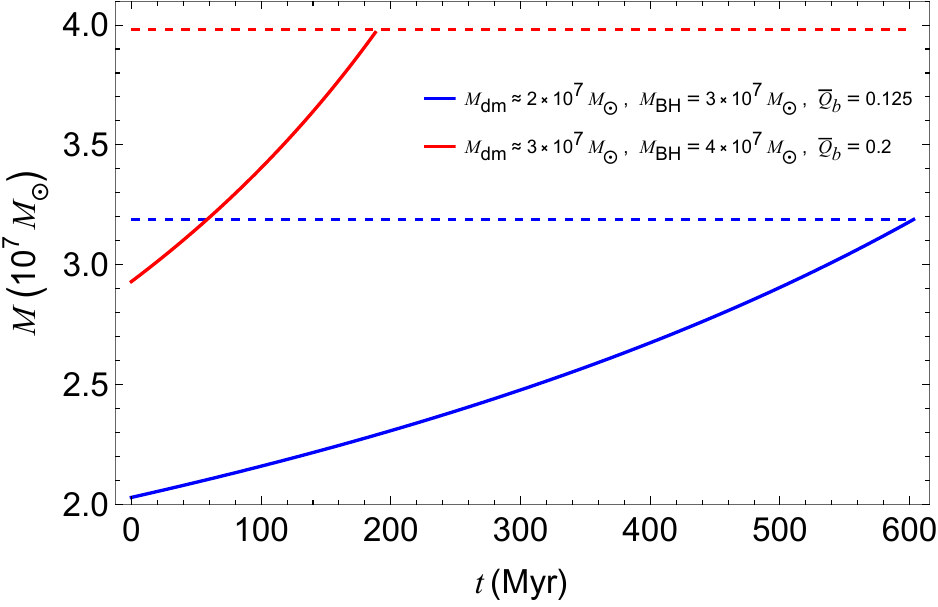}
    \caption{{Left: Time for SMBH formation by baryon-induced collapse as a function of $\bar{Q}_b$, for two selected examples of initial pure-dark-matter cores: $M_{\rm dm} = 2.03 \times 10^7 M_\odot$ (blue curve) and $M_{\rm dm}= 2.93 \times 10^7 M_\odot$ (red curve) which reach, respectively, the critical mass $M_{\rm BH} = 3.19\times 10^7 M_\odot$ (dot in Fig. \ref{fig:mrhoplane}) and $M_{\rm BH} = 3.98\times 10^7 M_\odot$. The gray-filled region shows the elapsed cosmological time from an initial redshift of $12\leq z_0\leq 15$ to a final redshift $z_f=10.3$, relevant for the analysis of UHZ1$^*$. Likewise, the green-filled region from $z_0=15$ to $4\leq z_0\leq 6$ is relevant for analyzing the so-called \textit{little red dots}. Right: Time evolution of the dark matter core mass while accreting baryonic matter, given by Eq. (\ref{eq:Mvst}), for the two examples in the left panel, for the values of $\bar{Q}_b$ in the legend. The dashed lines indicate the SMBH mass formed. Details in sections \ref{sec:3} and \ref{sec:4}.}}
    \label{fig:tBH}
\end{figure*}

Figure \ref{fig:tBH} shows $t_{\rm BH}$ for {two selected examples which, as we show in the next section, apply to two relevant astrophysical scenarios: the observed merging process of TXS 2116--077 with another nearby galaxy, and the farthest quasar recently observed by the \textit{Chandra} X-ray satellite, located in the background galaxy UHZ1 at $z= 10.3$ observed by the \textit{JWST}. In these two typical examples, the dark matter core is given, correspondingly, by the black-dashed curve shown in Fig. \ref{fig:mrhoplane} which has an initial mass $M_{\rm dm} \approx 2 \times 10^7 M_\odot$ (blue curve in Fig. \ref{fig:tBH}), and by another core having $M_{\rm dm} \approx 3 \times 10^7 M_\odot$ (red curve in Fig. \ref{fig:tBH}). The time to BH formation is plotted as a function of $\bar{Q}_b$,} for a range of astrophysically relevant values, e.g., as given by cosmological hydrodynamical simulations of the innermost gas density and velocity in high-$z$ halos \citep{2006MNRAS.368....2D,2022Natur.607...48L}, as well as in giant molecular clouds and clumps in the Milky Way, local starbursts, and distant galaxies \citep[see, e.g.,][]{2023MNRAS.519.6222D,2019NatAs...3.1115D,2017ApJ...834...57M}. {To exemplify, consider the initially stable core of $M_{\rm dm} = 2.03 \times 10^7 M_\odot$ (blue curve}, accreting baryonic gas with $\bar{\rho}_b = M_\odot$ pc$^{-3}$ and $v_b = 200$ km s$^{-1}$ \citep{2006MNRAS.368....2D}{, which leads to $\bar{Q}_b=0.125$.} The core attracts baryonic matter of such speed at $r = R_{\rm cap} = 4.36$ pc. {From Eqs. (\ref{eq:tau})--(\ref{eq:tcrit}),} the baryon-induced collapse occurs at $t_{\rm BH}\approx 0.6$ Gyr, forming an SMBH of $3.19\times 10^7 M_\odot$. Initially, the accretion rate is $\dot{M_b}= 0.012\,M_\odot$ yr$^{-1}$, {leading to an accretion luminosity below the Eddington limit, i.e.,}
{
\begin{align}
    L &= \beta \dot{M_b} c^2 = \beta\,6.81 \times 10^{44}\,\,\text{erg s}^{-1} < L_{\rm Edd}, \label{eq:MdotEdd}\\
    L_{\rm Edd} &=\frac{4 \pi G m_p c}{\sigma_T} M = 1.25\times 10^{45} \left(\frac{M}{10^7 M_\odot}\right) \,\text{erg s}^{-1},\label{eq:MdotEdd2}
\end{align}
}
where $\sigma_T$ is the Thomson scattering cross-section, $m_p$ is the proton mass, and $\beta$ is the efficiency in converting gravitational energy gain into electromagnetic radiation, {which is below unity. The ratio $L/L_{\rm Edd}$} is proportional to $M$, so the maximum value attained by this ratio is obtained for $M_{\rm crit}/M_{\rm dm}$, which is of order unity (in the present example, $1.57$). The accretion is sub-Eddington during the evolution to the critical mass point.

To be cautious, we have explored the occurrence of the baryon-induced collapse for different values of the baryon-to-dark-matter mass ratio, $\chi$, limiting ourselves to a maximum of $\chi_{\rm max} \approx \eta_{\rm max} = 0.8$; see Fig. \ref{fig:mrhoplane}. It is then clear from that figure that, given the maximum baryon-to-dark-matter mass ratio, there is a minimum constant-$N_{\rm dm}$ evolution (i.e. minimum $M_{\rm dm}$) below which no baryon-induced collapse occurs: the evolution track never crosses the red curve but a stable point of the $\eta_{\rm max} = 0.8$ sequence. Thus, we can estimate such a minimum mass of the initial pure-dark-matter core, $M^{(min)}_{\rm dm}$, for the occurrence of the baryon-induced collapse. Again, that value is a function of the given $\chi_{\rm max}$. We can readily obtain $M^{(\rm min)}_{\rm dm}$ by evaluating Eq. (\ref{eq:Mvst}) at the critical point for the maximum value of $\chi$, i.e.,
\begin{equation}\label{eq:Mdmmin}
    M^{\rm (min)}_{\rm dm} = \frac{M_{\rm BH}(\chi_{\rm max})}{1 + \chi_{\rm max}} \approx 0.22 M^{(0)}_{\rm crit},
\end{equation}
where, in the last expression, we have used Eq. (\ref{eq:Mcritchi}) with $\chi_{\rm max} \approx 0.8$. Therefore, for the fermion mass $m =100$ kev/$c^2$, we obtain $M^{\rm (min)}_{\rm dm} \approx 1.41 \times 10^7 M_\odot$. 

Equation (\ref{eq:Mdmmin}) tells that independently of $m$, the maximum ratio $\chi_{\rm max}$ sets a minimum value of the ratio $\mu$, i.e., $\mu_{\rm min} \equiv M^{\rm (min)}_{\rm dm}/M^{(0)}_{\rm crit} \approx 0.22$. From Eq. (\ref{eq:chiBH}), it turns out that the maximum dimensionless BH formation time is set by $\mu= \mu_{\rm min}$, i.e., $\bar{t}^{\rm max}_{\rm BH} \equiv \bar{t}_{\rm BH}(\mu_{\rm min}) \approx 0.44$. Therefore, the maximum BH formation time is $t^{\rm max}_{\rm BH} = 0.44\,\tau_{\rm max}$, where $\tau_{\rm max}$ is the value of $\tau$ for $M^{\rm (min)}_{\rm dm}$. By equating $t^{\rm max}_{\rm BH}$ to the universe lifetime ($\approx 13.8$ Gyr), we obtain an upper bound to $m$ over which the present baryon-induced collapse mechanism can form no BH:
\begin{equation}\label{eq:mmax}
    m_{\rm max} = 323.33 \sqrt{\frac{\bar{Q}_b}{0.1}}\,\,{\rm keV}/c^2.
\end{equation}
Although the upper limit (\ref{eq:mmax}) depends on $Q_b$ and has been obtained for the minimum pure-dark-matter core mass (which maximizes the BH formation time), it is remarkably similar to the upper bound of $350$ keV/$c^2$ obtained from a different theoretical and observational request: that the dark matter halo explains the Milky Way rotation curves \citep{2018PDU....21...82A} and that the core explains the orbits of the S-cluster stars in alternative to the central BH scenario for Sgr A* \citep{2020A&A...641A..34B, 2021MNRAS.505L..64B}. 

{\section{Astrophysical applications of the baryon-induced collapse}\label{sec:4}}

The SMBH formation by {the baryon-induced collapse of dark matter fermion cores} can trigger the activity at the center of galaxies by forming active galactic nuclei from merging events. This new scenario supports the hypothesis that relativistic jets in radio-loud active galactic nuclei are triggered by galaxy mergers  \citep[see, e.g.,][]{2015ApJ...806..147C}. An interesting case is the Seyfert galaxy TXS 2116--077, which is merging with another nearby Seyfert galaxy on an estimated timescale of $0.5$--$2$ Gyr. It hosts a radio jet with a kinematic age shorter than $15$ kyr, considerably shorter than the merger timescale, strengthening the above hypothesis \citep{2020ApJ...892..133P}. The derived mass of the central SMBH of TXS 2116--077 is $M_{\rm BH}\approx 3\times 10^7 M_\odot$, which agrees with the example presented {in Fig. \ref{fig:mrhoplane} (endpoint of black-dashed curve; blue curve in Fig. \ref{fig:tBH})}. The initial dark matter core of $M_{\rm dm} \approx 2\times 10^7 M_\odot$, accreting baryonic matter of density $\rho_b = 1\,M_\odot$ pc$^{-3}$ moving at $200$ km s$^{-1}$, so {$\bar{Q}_b = 0.125$}, collapses forming such an SMBH in $t_{\rm BH}\approx 0.6$ Gyr{; see Eqs. (\ref{eq:tau})--(\ref{eq:tcrit}) and Fig. \ref{fig:tBH}. These baryonic matter conditions agree with the TXS 2116--077 inner region observations \citep{2020ApJ...892..133P}, for a Compton thick active galactic nucleus with typical column density $N_H=10^{24}$ cm$^{-2}$}. The SMBH formation timescale increases (decreases) with decreasing (increasing) $Q_b$. It is worth recalling that the present theory predicts the core is surrounded by a dark matter halo of $M_h \sim 10^{11} M_\odot$ (see Fig. \ref{fig:profile}), in agreement with the TXS 2116--077 total mass \citep{2020ApJ...892..133P}. 

{Another case worth to discuss is the farthest quasar ever detected, observed in the X-rays by the \textit{Chandra} X-ray satellite, with a bolometric luminosity $L_{\rm bol} \approx 5\times 10^{45}$ erg s$^{-1}$ and associated SMBH mass $M_{\rm BH} \sim 10^7$--$10^8 M_\odot$, hosted by the \textit{JWST}-detected lensed galaxy UHZ1 at $z\approx 10.3$  \citep{2023NatAs.tmp..223B}. Assuming the SMBH accretes at the Eddington-limit rate, i.e., $L_{\rm bol} = L_{\rm Edd}$, we obtain from Eq. (\ref{eq:MdotEdd}), $M_{\rm BH}\approx 4\times 10^7 M_\odot\approx 0.63 M^{(0)}_{\rm crit}$. Thus, from Eq. (\ref{eq:Mcritchi}), $\chi_{\rm BH} \approx 0.35$, which via Eq. (\ref{eq:tcrit}) leads to $\bar{t}_{\rm BH} \approx 0.26$, so the initial mass of the dark matter core is $M_{\rm dm} = 0.74\,M_{\rm BH} \approx 3\times 10^7 M_\odot$. The red curve in Fig. \ref{fig:tBH} shows this case, i.e., $t_{\rm BH} = 0.26\,\tau$, where $\tau$ is given by Eq. (\ref{eq:tau}). The gray-filled region shows the cosmological time elapsed from an initial redshift $12\leq z_0\leq 15$ to the UHZ1 redshift, $z_f = 10.3$, which leads to $87.36\,\text{Myr}\lesssim t_{\rm BH}\lesssim 187.32\,\text{Myr}$. At fixed $z_f$, the larger the dark matter core seed redshift, $z_0$, the larger the time it has to accrete the necessary baryonic mass to trigger the collapse, so the smaller the $Q_b$. The cuts of the gray-filled region with the red curve give this example's corresponding range of solutions, i.e., $0.20 \lesssim \bar{Q}_b \lesssim 0.43$. This range suggests $\rho_b$ in UHZ1 about twice that of TXS 2116-077 (for similar $v_b$), in line with the column densities of \citet{2023NatAs.tmp..223B}.}  

{An SMBH of a larger mass than the above value would imply that the dark matter core collapsed at $z> 10.3$ and then continued to accrete until reaching that mass at $z_f=10.3$. Let us assume the SMBH has the largest inferred value from the observations, $M_{\rm BH}(z_f = 10.3) = 10^8 M_\odot$. The dark matter core collapsed at $z_c >z_f$, forming an SMBH of mass $M_{\rm BH}(z_c) = 4\times 10^7 M_\odot$. Thus, from $z_c$ to $z_f$, the newborn SMBH accreted $6 \times 10^7 M_\odot$. The SMBH mass $M_{\rm BH}(z_f)$ implies $L_{\rm bol}/L_{\rm Edd} = 0.4$. Assuming this ratio holds constant at this value, the time it takes for the BH to increase its mass from $M_{\rm BH}(z_c)$ to $M_{\rm BH}(z_f)$ is $\Delta t = 59.26$ Myr, for a gravitational-to-electromagnetic energy conversion factor $\beta = 0.057$, set by the binding energy of the last stable circular orbit for a Schwarzschild BH. Thus, the dark matter core collapsed at $z_c \approx 11.39$. If the dark matter \textit{core}--\textit{halo} distribution formed, e.g., at $z_0 = 15$, the baryon-induced collapse of the dark matter core must occur in $t_{\rm BH} = 127.88$ Myr, achievable for $\bar{Q}_b \approx 0.29$ (see Fig. \ref{fig:tBH}).}

{Therefore, the baryon-induced collapse of fermion dark matter cores explains the formation of SMBHs at high $z$. Consequently, it can also explain the \textit{little red dots}, i.e, the SMBHs of $10^7$--$10^8 M_\odot$ at $z\approx 4$--$6$ observed by the \textit{JWST} \citep[see, e.g.,][]{2023arXiv230605448M,2023ApJ...954L...4K}, for $\bar{Q}_b\sim 0.1$ (see Fig. \ref{fig:tBH}). Interestingly, the mean value of the \textit{little red dots} bolometric luminosity is $L_{\rm bol}/L_{\rm Edd} \approx 0.2$, similar to our previous example. Future studies can infer the SMBH mass function of the baryon-induced collapse scenario and compare it with the observed SMBH population at various redshifts.}
\\
\section{Conclusions}\label{sec:5}

How SMBHs form and grow at high cosmological redshifts has remained elusive for years \citep{2012Sci...337..544V, 2019PASA...36...27W}. The relevance of getting a satisfactory answer is exponentially increasing with the advent of the \textit{JWST}'s new, deep observations of the universe spacetime unveiling the farthest quasars ever detected \citep{2021ApJ...923..262Y,2022A&A...666A..17G,2023arXiv230801230M,2023ApJ...954L...4K,2023arXiv230904614Y,2023ARA&A..61..373F}. 

In \citet{2023MNRAS.523.2209A}, it was shown that the collapse of dark matter cores made of $50$--$100$ keV/$c^2$ fermions would lead to heavy BH seeds of $10^7$--$10^8 M_\odot$, which could grow further by sub-Eddington accretion of baryonic matter to reach $10^9 M_\odot$ in comfortable timescales of the order of a Gyr or less. This Letter answered the crucial questions of how a dark matter core reaches gravitational collapse conditions, how long that process could take, and how those conditions are realized in astrophysics and cosmology.

We have shown that an initially stable dark matter core, with a mass larger or equal to $22\%$ of the critical mass of a pure-dark-matter core, i.e., without ordinary/baryonic matter, can reach a point of gravitational collapse by gaining some threshold amount of baryons. We calculate such a critical mass for SMBH formation as a function of the baryon-to-dark-matter mass ratio. We limited our analysis to a maximum ratio of $\sim 80\%$. For this maximum ratio, the dark matter+baryon core critical mass is $\sim 40\%$ lower than that of a pure-dark-matter core. For $m=100$ keV/$c^2$ fermions, this baryon-induced collapse process leads to an SMBH of $\sim 10^7 M_\odot$. We provided analytic formulas for the SMBH mass as a function of the baryon-to-dark-matter mass ratio and the minimum mass of the initial dark matter core for the baryon-induced collapse to occur.

The gravitational capture of baryons from the environment triggers the gravitational collapse when it accumulates a threshold amount of baryons that \textit{instabilizes} the core equilibrium. The needed accretion rate is sub-Eddington for baryonic densities ($\rho_b \sim M_\odot$ pc$^{-3}$) and velocities ($v_b \sim 100$ km s$^{-1}$) leading to $Q_b = \rho_b/v_b^3$ values typical in cosmological hydrodynamical simulations of high-$z$ halos \citep{2006MNRAS.368....2D} (see also \citealp{2022Natur.607...48L}) and observations of giant molecular clouds and clumps in the Milky Way, local starbursts, and distant galaxies \citealp{2017ApJ...834...57M,2019NatAs...3.1115D,2023MNRAS.519.6222D}. The above process forms $\sim 10^7 M_\odot$ SMBHs in timescales shorter than a Gyr, for $m=100$ keV/$c^2$ fermions.

The SMBH formation timescale increases with the square of the fermion mass, so we obtain an upper bound to the latter of $m_{\rm max} \approx 323$ keV/$c^2$ for the collapse to occur in timescales shorter than the universe age, for a typical value of $Q_b$. The upper limit depends on $Q_b$ as $m_{\rm max} \propto \sqrt{Q_b}$. Imposing shorter timescales for the SMBH formation leads to a more stringent value of the dark matter fermion mass upper bound (at fixed $Q_b$).

Therefore, the accretion of baryonic matter by dark matter cores in the high-$z$ universe can lead to SMBHs by baryon-induced collapse in a fraction of a Gyr, which can further grow up to $\sim 10^8-10^9 M_\odot$ as required by the observations of the farthest quasars \citep{2023MNRAS.523.2209A}. {We have shown the viability of the baryon-induced collapse mechanism in three relevant cases: the formation of the SMBH in the Seyfert galaxy TXS 2116--077 in the merger with a nearby galaxy, the farthest quasar ever observed, located at $z=10.3$ at the center of the galaxy UHZ1, and the \textit{little red dots} at $z\approx 4$--$6$.}

{
The assessment of the above results of SMBH formation, together with the complementary application of the RAR model on galactic dynamics and structure formation (see, e.g., \citealp{2023Univ....9..197A}), point to a neutral, massive, spin 1/2 fermion with rest mass-energy between the one of active neutrinos and the one of electrons, of about $100$ keV. Does this fermion fit with any known dark matter particle candidates? A natural dark matter candidate that could be associated with our fermion is the right-handed sterile neutrinos introduced in the minimum standard model extension, $\nu$MSM \citep{nuMSM}. Under this assumption, we studied an extension of the RAR model with fermion self-interactions via a dark-sector massive (axial) vector mediator \citep{amrr} and calculated the self-interaction cross-section via an electroweak-like treatment. The latter was constrained using the bullet cluster and X-ray NuSTAR data from the Milky Way's central parsec, assuming the sterile neutrinos decay channel into photons (X-rays) and light (active) neutrinos. This analysis allows a $\sim 100$ keV fermion mass \citep{2020PDU....3000699Y}. On the other hand, promising direct searches of dark matter in terrestrial laboratories, e.g., Xenon, Krypton, and Argon detectors, via the dark matter interactions with ordinary matter, electrons, and nucleons, via kinetic energy recoils or even target ionization have started to look for a dark matter fermion in the tens of keV range \citep[see, e.g.,][and references therein]{2020JHEP...12..194S,2020PhRvL.124r1301D,2021PhRvD.103c5001D,2022PhRvL.129p1805A,2022JHEP...05..071L,2022JHEP...05..191G,2023PhRvL.131e2501R,2023Natur.618...47P,2022PhRvL.129p1804Z,2023PhRvL.130c1802A,2023PhRvD.108h3030C,2023JPhG...50a3001A,2023PhRvL.131b1002S}. A mass-energy which we have inferred from combined astrophysical analyses.
}

{The baryon-induced collapse of dark matter fermion cores, by answering the long-standing question of how SMBHs form and grow in the high-$z$ universe, adds a piece to the possible role of a yet-unobserved massive fermion in the Universe. It opens a research window verifiable, especially with the \textit{JWST} and \textit{Euclid} data, to further constrain the fermionic dark matter hypothesis. These constraints and information from galactic dynamics strengthen the synergy between astrophysics and terrestrial laboratories of direct searches of a \textit{light} dark matter particle below MeV energies.}

\section*{Acknowledgements}

C.R.A. is supported by the CONICET of Argentina, the ANPCyT (grant PICT-2018-03743), and ICRANet.


\begin{thebibliography}{}
\expandafter\ifx\csname natexlab\endcsname\relax\def\natexlab#1{#1}\fi
\providecommand{\url}[1]{\href{#1}{#1}}
\providecommand{\dodoi}[1]{doi:~\href{http://doi.org/#1}{\nolinkurl{#1}}}
\providecommand{\doeprint}[1]{\href{http://ascl.net/#1}{\nolinkurl{http://ascl.net/#1}}}
\providecommand{\doarXiv}[1]{\href{https://arxiv.org/abs/#1}{\nolinkurl{https://arxiv.org/abs/#1}}}

\bibitem[{{Aalbers} {et~al.}(2023){Aalbers}, {AbdusSalam}, {Abe}, {Aerne}, {Agostini}, {Ahmed Maouloud}, {Akerib}, {Akimov}, {Akshat}, {Al Musalhi}, {Alder}, {Alsum}, {Althueser}, {Amarasinghe}, {Amaro}, {Ames}, {Anderson}, {Andrieu}, {Angelides}, {Angelino}, {Angevaare}, {Antochi}, {Ant{\'o}n Martin}, {Antunovic}, {Aprile}, {Ara{\'u}jo}, {Armstrong}, {Arneodo}, {Arthurs}, {Asadi}, {Baek}, {Bai}, {Bajpai}, {Baker}, {Balajthy}, {Balashov}, {Balzer}, {Bandyopadhyay}, {Bang}, {Barberio}, {Bargemann}, {Baudis}, {Bauer}, {Baur}, {Baxter}, {Baxter}, {Bazyk}, {Beattie}, {Behrens}, {Bell}, {Bellagamba}, {Beltrame}, {Benabderrahmane}, {Bernard}, {Bertone}, {Bhattacharjee}, {Bhatti}, {Biekert}, {Biesiadzinski}, {Binau}, {Biondi}, {Biondi}, {Birch}, {Bishara}, {Bismark}, {Blanco}, {Blockinger}, {Bodnia}, {Boehm}, {Bolozdynya}, {Bolton}, {Bottaro}, {Bourgeois}, {Boxer}, {Br{\'a}s}, {Breskin}, {Breur}, {Brew}, {Brod}, {Brookes}, {Brown}, {Brown}, {Bruenner}, {Bruno}, {Budnik}, {Bui}, {Burdin}, {Buse}, {Busenitz},
  {Buttazzo}, {Buuck}, {Buzulutskov}, {Cabrita}, {Cai}, {Cai}, {Capelli}, {Cardoso}, {Carmona-Benitez}, {Cascella}, {Catena}, {Chakraborty}, {Chan}, {Chang}, {Chauvin}, {Chawla}, {Chen}, {Chepel}, {Chott}, {Cichon}, {Cimental Chavez}, {Cimmino}, {Clark}, {Co}, {Colijn}, {Conrad}, {Converse}, {Costa}, {Cottle}, {Cox}, {Creaner}, {Cuenca Garcia}, {Cussonneau}, {Cutter}, {Dahl}, {D'Andrea}, {David}, {Decowski}, {Dent}, {Deppisch}, {de Viveiros}, {Di Gangi}, {Di Giovanni}, {Di Pede}, {Dierle}, {Diglio}, {Dobson}, {Doerenkamp}, {Douillet}, {Drexlin}, {Druszkiewicz}, {Dunsky}, {Eitel}, {Elykov}, {Emken}, {Engel}, {Eriksen}, {Fairbairn}, {Fan}, {Fan}, {Farrell}, {Fayer}, {Fearon}, {Ferella}, {Ferrari}, {Fieguth}, {Fieguth}, {Fiorucci}, {Fischer}, {Flaecher}, {Flierman}, {Florek}, {Foot}, {Fox}, {Franceschini}, {Fraser}, {Frenk}, {Frohlich}, {Fruth}, {Fulgione}, {Fuselli}, {Gaemers}, {Gaior}, {Gaitskell}, {Galloway}, {Gao}, {Garcia Garcia}, {Genovesi}, {Ghag}, {Ghosh}, {Gibson}, {Gil}, {Giovagnoli}, {Girard},
  {Glade-Beucke}, {Gl{\"u}ck}, {Gokhale}, {de Gouv{\^e}a}, {Gr{\'a}f}, {Grandi}, {Grigat}, {Grinstein}, {van der Grinten}, {Gr{\"o}ssle}, {Guan}, {Guida}, {Gumbsheimer}, {Gwilliam}, {Hall}, {Hall}, {Hammann}, {Han}, {Hannen}, {Hansmann-Menzemer}, {Harata}, {Hardin}, {Hardy}, {Hardy}, {Harigaya}, {Harnik}, {Haselschwardt}, {Hernandez}, {Hertel}, {Higuera}, {Hils}, {Hochrein}, {Hoetzsch}, {Hoferichter}, {Hood}, {Hooper}, {Horn}, {Howlett}, {Huang}, {Huang}, {Hunt}, {Iacovacci}, {Iaquaniello}, {Ide}, {Ignarra}, {Iloglu}, {Itow}, {Jacquet}, {Jahangir}, {Jakob}, {James}, {Jansen}, {Ji}, {Ji}, {Joerg}, {Johnson}, {Joy}, {Kaboth}, {Kalhor}, {Kamaha}, {Kanezaki}, {Kar}, {Kara}, {Kato}, {Kavrigin}, {Kazama}, {Keaveney}, {Kellerer}, {Khaitan}, {Khazov}, {Khundzakishvili}, {Khurana}, {Kilminster}, {Kleifges}, {Ko}, {Kobayashi}, {Kodroff}, {Koltmann}, {Kopec}, {Kopmann}, {Kopp}, {Korley}, {Kornoukhov}, {Korolkova}, {Kraus}, {Krauss}, {Kravitz}, {Kreczko}, {Kudryavtsev}, {Kuger}, {Kumar}, {L{\'o}pez Paredes}, {LaCascio},
  {Laha}, {Laine}, {Landsman}, {Lang}, {Leason}, {Lee}, {Leonard}, {Lesko}, {Levinson}, {Levy}, {Li}, {Li}, {Li}, {Liang}, {Liebenthal}, {Lin}, {Lin}, {Lindemann}, {Lindner}, {Lindote}, {Linehan}, {Lippincott}, {Liu}, {Liu}, {Liu}, {Loizeau}, {Lombardi}, {Long}, {Lopes}, {Lopez Asamar}, {Lorenzon}, {Lu}, {Luitz}, {Ma}, {Machado}, {Macolino}, {Maeda}, {Mahlstedt}, {Majewski}, {Manalaysay}, {Mancuso}, {Manenti}, {Manfredini}, {Mannino}, {Marangou}, {March-Russell}, {Marignetti}, {Marrod{\'a}n Undagoitia}, {Martens}, {Martin}, {Martinez-Soler}, {Masbou}, {Masson}, {Masson}, {Mastroianni}, {Mastronardi}, {Matias-Lopes}, {McCarthy}, {McFadden}, {McGinness}, {McKinsey}, {McLaughlin}, {McMichael}, {Meinhardt}, {Men{\'e}ndez}, {Meng}, {Messina}, {Midha}, {Milisavljevic}, {Miller}, {Milosevic}, {Milutinovic}, {Mitra}, {Miuchi}, {Mizrachi}, {Mizukoshi}, {Molinario}, {Monte}, {Monteiro}, {Monzani}, {Moore}, {Mor{\r{a}}}, {Morad}, {Morales Mendoza}, {Moriyama}, {Morrison}, {Morteau}, {Mosbacher}, {Mount}, {Mueller},
  {Murphy}, {Murra}, {Naim}, {Nakamura}, {Nash}, {Navaieelavasani}, {Naylor}, {Nedlik}, {Nelson}, {Neves}, {Newstead}, {Ni}, {Nikoleyczik}, {Niro}, {Oberlack}, {Obradovic}, {Odgers}, {O'Hare}, {Oikonomou}, {Olcina}, {Oliver-Mallory}, {Oranday}, {Orpwood}, {Ostrovskiy}, {Ozaki}, {Paetsch}, {Pal}, {Palacio}, {Palladino}, {Palmer}, {Panci}, {Pandurovic}, {Parlati}, {Parveen}, {Patton}, {P{\v{e}}{\v{c}}}, {Pellegrini}, {Penning}, {Pereira}, {Peres}, {Perez-Gonzalez}, {Perry}, {Pershing}, {Petrossian-Byrne}, {Pienaar}, {Piepke}, {Pieramico}, {Pierre}, {Piotter}, {Pizzella}, {Plante}, {Pollmann}, {Porzio}, {Qi}, {Qie}, {Qin}, {Quevedo}, {Raj}, {Rajado Silva}, {Ramanathan}, {Ram{\'\i}rez Garc{\'\i}a}, {Ravanis}, {Redard-Jacot}, {Redigolo}, {Reichard}, {Reichenbacher}, {Rhyne}, {Richards}, {Riffard}, {Rischbieter}, {Rocchetti}, {Rosenfeld}, {Rosero}, {Rupp}, {Rushton}, {Saha}, {Salucci}, {Sanchez}, {Sanchez-Lucas}, {Santone}, {dos Santos}, {Sarnoff}, {Sartorelli}, {Sazzad}, {Scheibelhut}, {Schnee}, {Schrank},
  {Schreiner}, {Schulte}, {Schulte}, {Schulze Eissing}, {Schumann}, {Schwemberger}, {Schwenk}, {Schwetz}, {Scotto Lavina}, {Scovell}, {Sekiya}, {Selvi}, {Semenov}, {Semeria}, {Shagin}, {Shaw}, {Shi}, {Shockley}, {Shutt}, {Si-Ahmed}, {Silk}, {Silva}, {Silva}, {Simgen}, {{\v{S}}imkovic}, {Sinev}, {Singh}, {Skulski}, {Smirnov}, {Smith}, {Solmaz}, {Solovov}, {Sorensen}, {Soria}, {Sparmann}, {Stancu}, {Steidl}, {Stevens}, {Stifter}, {Strigari}, {Subotic}, {Suerfu}, {Suliga}, {Sumner}, {Szabo}, {Szydagis}, {Takeda}, {Takeuchi}, {Tan}, {Taricco}, {Taylor}, {Temples}, {Terliuk}, {Terman}, {Thers}, {Thieme}, {Th{\"u}mmler}, {Tiedt}, {Timalsina}, {To}, {Toennies}, {Tong}, {Toschi}, {Tovey}, {Tranter}, {Trask}, {Trinchero}, {Tripathi}, {Tronstad}, {Trotta}, {Tsai}, {Tunnell}, {Turner}, {Ueno}, {Urquijo}, {Utku}, {Vaitkus}, {Valerius}, {Vassilev}, {Vecchi}, {Velan}, {Vetter}, {Vincent}, {Vittorio}, {Volta}, {von Krosigk}, {von Piechowski}, {Vorkapic}, {Wagner}, {Wang}, {Wang}, {Wang}, {Wang}, {Wang}, {Wang}, {Wang},
  {Wang}, {Watson}, {Wei}, {Weinheimer}, {Weisman}, {Weiss}, {Wenz}, {West}, {Whitis}, {Williams}, {Wilson}, {Winkler}, {Wittweg}, {Wolf}, {Wolf}, {Wolfs}, {Woodford}, {Woodward}, {Wright}, {Wu}, {Wu}, {W{\"u}stling}, {Wurm}, {Xia}, {Xiang}, {Xing}, {Xu}, {Xu}, {Xu}, {Yamashita}, {Yamazaki}, {Yan}, {Yang}, {Yang}, {Ye}, {Yeh}, {Young}, {Yu}, {Yu}, {Yuan}, {Zavattini}, {Zerbo}, {Zhang}, {Zhong}, {Zhou}, {Zhou}, {Zhu}, {Zhu}, {Zhuang}, {Zopounidis}, {Zuber}, \& {Zupan}}]{2023JPhG...50a3001A}
{Aalbers}, J., {AbdusSalam}, S.~S., {Abe}, K., {et~al.} 2023, Journal of Physics G Nuclear Physics, 50, 013001, \dodoi{10.1088/1361-6471/ac841a}

\bibitem[{{Abe} {et~al.}(2023){Abe}, {Hayato}, {Hiraide}, {Ieki}, {Ikeda}, {Kameda}, {Kanemura}, {Kaneshima}, {Kashiwagi}, {Kataoka}, {Miki}, {Mine}, {Miura}, {Moriyama}, {Nakano}, {Nakahata}, {Nakayama}, {Noguchi}, {Okamoto}, {Sato}, {Sekiya}, {Shiba}, {Shimizu}, {Shiozawa}, {Sonoda}, {Suzuki}, {Takeda}, {Takemoto}, {Takenaka}, {Tanaka}, {Watanabe}, {Yano}, {Han}, {Kajita}, {Okumura}, {Tashiro}, {Tomiya}, {Wang}, {Xia}, {Yoshida}, {Megias}, {Fernandez}, {Labarga}, {Ospina}, {Zaldivar}, {Pointon}, {Kearns}, {Raaf}, {Wan}, {Wester}, {Bian}, {Griskevich}, {Kropp}, {Locke}, {Smy}, {Sobel}, {Takhistov}, {Yankelevich}, {Hill}, {Park}, {Bodur}, {Scholberg}, {Walter}, {Bernard}, {Coffani}, {Drapier}, {El Hedri}, {Giampaolo}, {Mueller}, {Santos}, {Paganini}, {Quilain}, {Ishizuka}, {Nakamura}, {Jang}, {Learned}, {Choi}, {Cao}, {Anthony}, {Martin}, {Scott}, {Sztuc}, {Uchida}, {Berardi}, {Catanesi}, {Radicioni}, {Calabria}, {Machado}, {De Rosa}, {Collazuol}, {Iacob}, {Lamoureux}, {Mattiazzi}, {Ludovici}, {Gonin},
  {Pronost}, {Fujisawa}, {Maekawa}, {Nishimura}, {Friend}, {Hasegawa}, {Ishida}, {Kobayashi}, {Jakkapu}, {Matsubara}, {Nakadaira}, {Nakamura}, {Oyama}, {Sakashita}, {Sekiguchi}, {Tsukamoto}, {Boschi}, {Di Lodovico}, {Gao}, {Goldsack}, {Katori}, {Migenda}, {Taani}, {Zsoldos}, {Kotsar}, {Ozaki}, {Suzuki}, {Takeuchi}, {Bronner}, {Feng}, {Kikawa}, {Mori}, {Nakaya}, {Wendell}, {Yasutome}, {Jenkins}, {McCauley}, {Mehta}, {Tsui}, {Fukuda}, {Itow}, {Menjo}, {Ninomiya}, {Lagoda}, {Lakshmi}, {Mandal}, {Mijakowski}, {Prabhu}, {Zalipska}, {Jia}, {Jiang}, {Jung}, {Wilking}, {Yanagisawa}, {Harada}, {Ishino}, {Ito}, {Kitagawa}, {Koshio}, {Nakanishi}, {Sakai}, {Barr}, {Barrow}, {Cook}, {Samani}, {Wark}, {Nova}, {Yang}, {Malek}, {McElwee}, {Stone}, {Thiesse}, {Thompson}, {Okazawa}, {Kim}, {Seo}, {Yu}, {Ichikawa}, {Nakamura}, {Tairafune}, {Nishijima}, {Iwamoto}, {Nakagiri}, {Nakajima}, {Taniuchi}, {Yokoyama}, {Martens}, {de Perio}, {Vagins}, {Kuze}, {Izumiyama}, {Inomoto}, {Ishitsuka}, {Ito}, {Kinoshita}, {Matsumoto},
  {Ommura}, {Shigeta}, {Shinoki}, {Suganuma}, {Yamauchi}, {Martin}, {Tanaka}, {Towstego}, {Akutsu}, {Gousy-Leblanc}, {Hartz}, {Konaka}, {Prouse}, {Chen}, {Xu}, {Zhang}, {Posiadala-Zezula}, {Hadley}, {Nicholson}, {O'Flaherty}, {Richards}, {Ali}, {Jamieson}, {Marti}, {Minamino}, {Pintaudi}, {Sano}, {Suzuki}, {Wada}, \& {Super-Kamiokande Collaboration}}]{2023PhRvL.130c1802A}
{Abe}, K., {Hayato}, Y., {Hiraide}, K., {et~al.} 2023, \prl, 130, 031802, \dodoi{10.1103/PhysRevLett.130.031802}

\bibitem[{{Aprile} {et~al.}(2022){Aprile}, {Abe}, {Agostini}, {Ahmed Maouloud}, {Althueser}, {Andrieu}, {Angelino}, {Angevaare}, {Antochi}, {Ant{\'o}n Martin}, {Arneodo}, {Baudis}, {Baxter}, {Bellagamba}, {Biondi}, {Bismark}, {Brown}, {Bruenner}, {Bruno}, {Budnik}, {Bui}, {Cai}, {Capelli}, {Cardoso}, {Cichon}, {Clark}, {Colijn}, {Conrad}, {Cuenca-Garc{\'\i}a}, {Cussonneau}, {D'Andrea}, {Decowski}, {di Gangi}, {di Pede}, {di Giovanni}, {di Stefano}, {Diglio}, {Eitel}, {Elykov}, {Farrell}, {Ferella}, {Ferrari}, {Fischer}, {Fulgione}, {Gaemers}, {Gaior}, {Gallo Rosso}, {Galloway}, {Gao}, {Gardner}, {Glade-Beucke}, {Grandi}, {Grigat}, {Guida}, {Hammann}, {Higuera}, {Hils}, {Hoetzsch}, {Howlett}, {Iacovacci}, {Itow}, {Jakob}, {Joerg}, {Joy}, {Kato}, {Kara}, {Kavrigin}, {Kazama}, {Kobayashi}, {Koltman}, {Kopec}, {Kuger}, {Landsman}, {Lang}, {Levinson}, {Li}, {Li}, {Liang}, {Lindemann}, {Lindner}, {Liu}, {Loizeau}, {Lombardi}, {Long}, {Lopes}, {Ma}, {Macolino}, {Mahlstedt}, {Mancuso}, {Manenti}, {Marignetti},
  {Marrod{\'a}n Undagoitia}, {Martens}, {Masbou}, {Masson}, {Masson}, {Mastroianni}, {Messina}, {Miuchi}, {Mizukoshi}, {Molinario}, {Moriyama}, {Mor{\^a}}, {Mosbacher}, {Murra}, {M{\"u}ller}, {Ni}, {Oberlack}, {Paetsch}, {Palacio}, {Paschos}, {Peres}, {Peters}, {Pienaar}, {Pierre}, {Pizzella}, {Plante}, {Qi}, {Qin}, {Ram{\'\i}rez Garc{\'\i}a}, {Reichard}, {Rocchetti}, {Rupp}, {Sanchez}, {Dos Santos}, {Sarnoff}, {Sartorelli}, {Schreiner}, {Schulte}, {Schulte}, {Schulze Ei{\ss}ing}, {Schumann}, {Scotto Lavina}, {Selvi}, {Semeria}, {Shagin}, {Shi}, {Shockley}, {Silva}, {Simgen}, {Stephen}, {Takeda}, {Tan}, {Terliuk}, {Thers}, {Toschi}, {Trinchero}, {Tunnell}, {T{\"o}nnies}, {Valerius}, {Volta}, {Wei}, {Weinheimer}, {Weiss}, {Wenz}, {Wittweg}, {Wolf}, {Xu}, {Xu}, {Yamashita}, {Yang}, {Ye}, {Yuan}, {Zavattini}, {Zhong}, {Zhu}, \& {Xenon Collaboration}}]{2022PhRvL.129p1805A}
{Aprile}, E., {Abe}, K., {Agostini}, F., {et~al.} 2022, \prl, 129, 161805, \dodoi{10.1103/PhysRevLett.129.161805}

\bibitem[{{Arg{\"u}elles} {et~al.}(2023{\natexlab{a}}){Arg{\"u}elles}, {Becerra-Vergara}, {Rueda}, \& {Ruffini}}]{2023Univ....9..197A}
{Arg{\"u}elles}, C.~R., {Becerra-Vergara}, E.~A., {Rueda}, J.~A., \& {Ruffini}, R. 2023{\natexlab{a}}, Universe, 9, 197, \dodoi{10.3390/universe9040197}

\bibitem[{{Arg{\"u}elles} {et~al.}(2023{\natexlab{b}}){Arg{\"u}elles}, {Boshkayev}, {Krut}, {Nurbakhyt}, {Rueda}, {Ruffini}, {Uribe-Su{\'a}rez}, \& {Yunis}}]{2023MNRAS.523.2209A}
{Arg{\"u}elles}, C.~R., {Boshkayev}, K., {Krut}, A., {et~al.} 2023{\natexlab{b}}, \mnras, 523, 2209, \dodoi{10.1093/mnras/stad1380}

\bibitem[{{Arg{\"u}elles} \& et~al.(2018)}]{2018PDU....21...82A}
{Arg{\"u}elles}, C.~R., \& et~al. 2018, Phys. Dark Universe, 21, 82, \dodoi{10.1016/j.dark.2018.07.002}

\bibitem[{{Arg{\"u}elles} \& et~al.(2019)}]{2019PDU....24..278A}
---. 2019, Phys. Dark Universe, 24, 100278, \dodoi{https://doi.org/10.1016/j.dark.2019.100278}

\bibitem[{{Arg{\"u}elles} \& et~al.(2021)}]{2021MNRAS.502.4227A}
---. 2021, \mnras, 502, 4227, \dodoi{10.1093/mnras/staa3986}

\bibitem[{{Arg{\"u}elles} {et~al.}(2016){Arg{\"u}elles}, Mavromatos, Rueda, \& Ruffini}]{amrr}
{Arg{\"u}elles}, C.~R., Mavromatos, N.~E., Rueda, J.~A., \& Ruffini, R. 2016, JCAP, 1604, 038, \dodoi{10.1088/1475-7516/2016/04/038}

\bibitem[{{Arg{\"u}elles} {et~al.}(2022){Arg{\"u}elles}, {Mestre}, {Becerra-Vergara}, {Crespi}, {Krut}, {Rueda}, \& {Ruffini}}]{2022MNRAS.511L..35A}
{Arg{\"u}elles}, C.~R., {Mestre}, M.~F., {Becerra-Vergara}, E.~A., {et~al.} 2022, \mnras, 511, L35, \dodoi{10.1093/mnrasl/slab126}

\bibitem[{{Arita} {et~al.}(2023){Arita}, {Kashikawa}, {Matsuoka}, {He}, {Ito}, {Liang}, {Ishimoto}, {Yoshioka}, {Takeda}, {Iwasawa}, {Onoue}, {Toba}, \& {Imanishi}}]{2023arXiv230702531A}
{Arita}, J., {Kashikawa}, N., {Matsuoka}, Y., {et~al.} 2023, arXiv e-prints, arXiv:2307.02531, \dodoi{10.48550/arXiv.2307.02531}

\bibitem[{{Becerra-Vergara} {et~al.}(2020){Becerra-Vergara}, {Arg{\"u}elles}, \& et~al.}]{2020A&A...641A..34B}
{Becerra-Vergara}, E.~A., {Arg{\"u}elles}, C.~R., \& et~al. 2020, \aap, 641, A34, \dodoi{10.1051/0004-6361/201935990}

\bibitem[{{Becerra-Vergara} \& et~al.(2021)}]{2021MNRAS.505L..64B}
{Becerra-Vergara}, E.~A., \& et~al. 2021, \mnras, 505, L64, \dodoi{10.1093/mnrasl/slab051}

\bibitem[{{Begelman} {et~al.}(2008){Begelman}, {Rossi}, \& {Armitage}}]{2008MNRAS.387.1649B}
{Begelman}, M.~C., {Rossi}, E.~M., \& {Armitage}, P.~J. 2008, \mnras, 387, 1649, \dodoi{10.1111/j.1365-2966.2008.13344.x}

\bibitem[{{Begelman} {et~al.}(2006){Begelman}, {Volonteri}, \& {Rees}}]{2006MNRAS.370..289B}
{Begelman}, M.~C., {Volonteri}, M., \& {Rees}, M.~J. 2006, \mnras, 370, 289, \dodoi{10.1111/j.1365-2966.2006.10467.x}

\bibitem[{{Bogdan} {et~al.}(2023){Bogdan}, {Goulding}, {Natarajan}, {Kovacs}, {Tremblay}, {Chadayammuri}, {Volonteri}, {Kraft}, {Forman}, {Jones}, {Churazov}, \& {Zhuravleva}}]{2023arXiv230515458B}
{Bogdan}, A., {Goulding}, A., {Natarajan}, P., {et~al.} 2023, arXiv e-prints, arXiv:2305.15458, \dodoi{10.48550/arXiv.2305.15458}

\bibitem[{{Bogd{\'a}n} {et~al.}(2023){Bogd{\'a}n}, {Goulding}, {Natarajan}, {Kov{\'a}cs}, {Tremblay}, {Chadayammuri}, {Volonteri}, {Kraft}, {Forman}, {Jones}, {Churazov}, \& {Zhuravleva}}]{2023NatAs.tmp..223B}
{Bogd{\'a}n}, {\'A}., {Goulding}, A.~D., {Natarajan}, P., {et~al.} 2023, Nature Astronomy, \dodoi{10.1038/s41550-023-02111-9}

\bibitem[{{Caddell} {et~al.}(2023){Caddell}, {Flambaum}, \& {Roberts}}]{2023PhRvD.108h3030C}
{Caddell}, A.~R., {Flambaum}, V.~V., \& {Roberts}, B.~M. 2023, \prd, 108, 083030, \dodoi{10.1103/PhysRevD.108.083030}

\bibitem[{{Chavanis}(1998)}]{1998MNRAS.300..981C}
{Chavanis}, P.-H. 1998, \mnras, 300, 981, \dodoi{10.1046/j.1365-8711.1998.01867.x}

\bibitem[{{Chavanis} {et~al.}(2015){Chavanis}, {Lemou}, \& {M{\'e}hats}}]{2015PhRvD..92l3527C}
{Chavanis}, P.-H., {Lemou}, M., \& {M{\'e}hats}, F. 2015, \prd, 92, 123527, \dodoi{10.1103/PhysRevD.92.123527}

\bibitem[{{Chiaberge} {et~al.}(2015){Chiaberge}, {Gilli}, {Lotz}, \& {Norman}}]{2015ApJ...806..147C}
{Chiaberge}, M., {Gilli}, R., {Lotz}, J.~M., \& {Norman}, C. 2015, \apj, 806, 147, \dodoi{10.1088/0004-637X/806/2/147}

\bibitem[{{Dekel} \& {Birnboim}(2006)}]{2006MNRAS.368....2D}
{Dekel}, A., \& {Birnboim}, Y. 2006, \mnras, 368, 2, \dodoi{10.1111/j.1365-2966.2006.10145.x}

\bibitem[{{Dessauges-Zavadsky} {et~al.}(2019){Dessauges-Zavadsky}, {Richard}, {Combes}, \& {et al.}}]{2019NatAs...3.1115D}
{Dessauges-Zavadsky}, M., {Richard}, J., {Combes}, F., \& {et al.} 2019, Nature Astronomy, 3, 1115, \dodoi{10.1038/s41550-019-0874-0}

\bibitem[{{Dessauges-Zavadsky} {et~al.}(2023){Dessauges-Zavadsky}, {Richard}, {Combes}, {Messa}, {Nagy}, {Mayer}, {Schaerer}, {Egami}, \& {Adamo}}]{2023MNRAS.519.6222D}
{Dessauges-Zavadsky}, M., {Richard}, J., {Combes}, F., {et~al.} 2023, \mnras, 519, 6222, \dodoi{10.1093/mnras/stad113}

\bibitem[{{Dror} {et~al.}(2020){Dror}, {Elor}, \& {McGehee}}]{2020PhRvL.124r1301D}
{Dror}, J.~A., {Elor}, G., \& {McGehee}, R. 2020, \prl, 124, 181301, \dodoi{10.1103/PhysRevLett.124.181301}

\bibitem[{{Dror} {et~al.}(2021){Dror}, {Elor}, {McGehee}, \& {Yu}}]{2021PhRvD.103c5001D}
{Dror}, J.~A., {Elor}, G., {McGehee}, R., \& {Yu}, T.-T. 2021, \prd, 103, 035001, \dodoi{10.1103/PhysRevD.103.035001}

\bibitem[{{Fan} {et~al.}(2023){Fan}, {Ba{\~n}ados}, \& {Simcoe}}]{2023ARA&A..61..373F}
{Fan}, X., {Ba{\~n}ados}, E., \& {Simcoe}, R.~A. 2023, \araa, 61, 373, \dodoi{10.1146/annurev-astro-052920-102455}

\bibitem[{{Ge} {et~al.}(2022){Ge}, {He}, {Ma}, \& {Sheng}}]{2022JHEP...05..191G}
{Ge}, S.-F., {He}, X.-G., {Ma}, X.-D., \& {Sheng}, J. 2022, Journal of High Energy Physics, 2022, 191, \dodoi{10.1007/JHEP05(2022)191}

\bibitem[{{Gilli} {et~al.}(2022){Gilli}, {Norman}, {Calura}, {Vito}, {Decarli}, {Marchesi}, {Iwasawa}, {Comastri}, {Lanzuisi}, {Pozzi}, {D'Amato}, {Vignali}, {Brusa}, {Mignoli}, \& {Cox}}]{2022A&A...666A..17G}
{Gilli}, R., {Norman}, C., {Calura}, F., {et~al.} 2022, \aap, 666, A17, \dodoi{10.1051/0004-6361/202243708}

\bibitem[{{Hosokawa} {et~al.}(2016){Hosokawa}, {Hirano}, {Kuiper}, {Yorke}, {Omukai}, \& {Yoshida}}]{2016ApJ...824..119H}
{Hosokawa}, T., {Hirano}, S., {Kuiper}, R., {et~al.} 2016, \apj, 824, 119, \dodoi{10.3847/0004-637X/824/2/119}

\bibitem[{{Inayoshi} {et~al.}(2020){Inayoshi}, {Visbal}, \& {Haiman}}]{2020ARA&A..58...27I}
{Inayoshi}, K., {Visbal}, E., \& {Haiman}, Z. 2020, \araa, 58, 27, \dodoi{10.1146/annurev-astro-120419-014455}

\bibitem[{{Kocevski} {et~al.}(2023){Kocevski}, {Onoue}, {Inayoshi}, \& {et al.}}]{2023ApJ...954L...4K}
{Kocevski}, D.~D., {Onoue}, M., {Inayoshi}, K., \& {et al.} 2023, \apjl, 954, L4, \dodoi{10.3847/2041-8213/ace5a0}

\bibitem[{{Krut} {et~al.}(2023){Krut}, {Arg{\"u}elles}, {Chavanis}, {Rueda}, \& {Ruffini}}]{2023ApJ...945....1K}
{Krut}, A., {Arg{\"u}elles}, C.~R., {Chavanis}, P.~H., {Rueda}, J.~A., \& {Ruffini}, R. 2023, \apj, 945, 1, \dodoi{10.3847/1538-4357/acb8bd}

\bibitem[{{Larson} {et~al.}(2023){Larson}, {Finkelstein}, {Kocevski}, \& {et al.}}]{2023ApJ...953L..29L}
{Larson}, R.~L., {Finkelstein}, S.~L., {Kocevski}, D.~D., \& {et al.} 2023, \apjl, 953, L29, \dodoi{10.3847/2041-8213/ace619}

\bibitem[{{Latif} {et~al.}(2022){Latif}, {Whalen}, {Khochfar}, {Herrington}, \& {Woods}}]{2022Natur.607...48L}
{Latif}, M.~A., {Whalen}, D.~J., {Khochfar}, S., {Herrington}, N.~P., \& {Woods}, T.~E. 2022, \nat, 607, 48, \dodoi{10.1038/s41586-022-04813-y}

\bibitem[{{Li} {et~al.}(2022){Li}, {Liao}, \& {Zhang}}]{2022JHEP...05..071L}
{Li}, T., {Liao}, J., \& {Zhang}, R.-J. 2022, Journal of High Energy Physics, 2022, 71, \dodoi{10.1007/JHEP05(2022)071}

\bibitem[{{Madau} \& {Rees}(2001)}]{2001ApJ...551L..27M}
{Madau}, P., \& {Rees}, M.~J. 2001, \apjl, 551, L27, \dodoi{10.1086/319848}

\bibitem[{{Maiolino} {et~al.}(2023){Maiolino}, {Scholtz}, {Curtis-Lake}, \& {et al.}}]{2023arXiv230801230M}
{Maiolino}, R., {Scholtz}, J., {Curtis-Lake}, E., \& {et al.} 2023, arXiv e-prints, arXiv:2308.01230, \dodoi{10.48550/arXiv.2308.01230}

\bibitem[{{Matthee} {et~al.}(2023){Matthee}, {Naidu}, {Brammer}, {Chisholm}, {Eilers}, {Goulding}, {Greene}, {Kashino}, {Labbe}, {Lilly}, {Mackenzie}, {Oesch}, {Weibel}, {Wuyts}, {Xiao}, {Bordoloi}, {Bouwens}, {van Dokkum}, {Illingworth}, {Kramarenko}, {Maseda}, {Mason}, {Meyer}, {Nelson}, {Reddy}, {Shivaei}, {Simcoe}, \& {Yue}}]{2023arXiv230605448M}
{Matthee}, J., {Naidu}, R.~P., {Brammer}, G., {et~al.} 2023, arXiv e-prints, arXiv:2306.05448, \dodoi{10.48550/arXiv.2306.05448}

\bibitem[{{Miville-Desch{\^e}nes} {et~al.}(2017){Miville-Desch{\^e}nes}, {Murray}, \& {Lee}}]{2017ApJ...834...57M}
{Miville-Desch{\^e}nes}, M.-A., {Murray}, N., \& {Lee}, E.~J. 2017, \apj, 834, 57, \dodoi{10.3847/1538-4357/834/1/57}

\bibitem[{{Oppenheimer} \& {Volkoff}(1939)}]{1939PhRv...55..374O}
{Oppenheimer}, J.~R., \& {Volkoff}, G.~M. 1939, Physical Review, 55, 374, \dodoi{10.1103/PhysRev.55.374}

\bibitem[{{Pacucci} {et~al.}(2023){Pacucci}, {Nguyen}, {Carniani}, {Maiolino}, \& {Fan}}]{2023arXiv230812331P}
{Pacucci}, F., {Nguyen}, B., {Carniani}, S., {Maiolino}, R., \& {Fan}, X. 2023, arXiv e-prints, arXiv:2308.12331, \dodoi{10.48550/arXiv.2308.12331}

\bibitem[{{Paliya} {et~al.}(2020){Paliya}, {P{\'e}rez}, {Garc{\'\i}a-Benito}, {et al.}, {Prada}, {Alberdi}, {Suh}, {Chandra}, {Dom{\'\i}nguez}, {Marchesi}, {Di Matteo}, {Hartmann}, \& {Chiaberge}}]{2020ApJ...892..133P}
{Paliya}, V.~S., {P{\'e}rez}, E., {Garc{\'\i}a-Benito}, R., {et~al.} 2020, \apj, 892, 133, \dodoi{10.3847/1538-4357/ab754f}

\bibitem[{{PandaX Collaboration} {et~al.}(2023){PandaX Collaboration}, {Abdukerim}, {Bo}, {Cui}, {Chen}, {Chen}, {Cheng}, {Cheng}, {Fan}, {Fang}, {Fu}, {Fu}, {Geng}, {Giboni}, {Gu}, {Guo}, {Han}, {Han}, {He}, {He}, {Huang}, {Huang}, {Huang}, {Huang}, {Hou}, {Hou}, {Ji}, {Ju}, {Li}, {Li}, {Li}, {Li}, {Li}, {Lin}, {Liu}, {Lu}, {Lu}, {Luo}, {Luo}, {Ma}, {Ma}, {Mao}, {Meng}, {Qi}, {Qian}, {Ren}, {Shaheed}, {Shang}, {Shao}, {Shen}, {Si}, {Sun}, {Tan}, {Tao}, {Wang}, {Wang}, {Wang}, {Wang}, {Wang}, {Wang}, {Wang}, {Wang}, {Wei}, {Wu}, {Wu}, {Xia}, {Xiao}, {Xiao}, {Xie}, {Yan}, {Yan}, {Yang}, {Yang}, {Yao}, {Yu}, {Yuan}, {Yuan}, {Zeng}, {Zhang}, {Zhang}, {Zhang}, {Zhang}, {Zhang}, {Zhang}, {Zhang}, {Zhang}, {Zhang}, {Zhang}, {Zhao}, {Zheng}, {Zhou}, {Zhou}, {Zhou}, {Zhou}, \& {Zhou}}]{2023Natur.618...47P}
{PandaX Collaboration}, Ning, X., {Abdukerim}, A., {Bo}, Z., {et~al.} 2023, \nat, 618, 47, \dodoi{10.1038/s41586-023-05982-0}

\bibitem[{{Rebeiro} {et~al.}(2023){Rebeiro}, {Triambak}, {Garrett}, {Ball}, {Brown}, {Men{\'e}ndez}, {Romeo}, {Adsley}, {Lenardo}, {Lindsay}, {Bildstein}, {Burbadge}, {Coleman}, {Diaz Varela}, {Dubey}, {Faestermann}, {Hertenberger}, {Kamil}, {Leach}, {Natzke}, {Nzobadila Ondze}, {Radich}, {Rand}, \& {Wirth}}]{2023PhRvL.131e2501R}
{Rebeiro}, B.~M., {Triambak}, S., {Garrett}, P.~E., {et~al.} 2023, \prl, 131, 052501, \dodoi{10.1103/PhysRevLett.131.052501}

\bibitem[{{Ruffini} {et~al.}(2015){Ruffini}, {Arg{\"u}elles}, \& {Rueda}}]{2015MNRAS.451..622R}
{Ruffini}, R., {Arg{\"u}elles}, C.~R., \& {Rueda}, J.~A. 2015, \mnras, 451, 622, \dodoi{10.1093/mnras/stv1016}

\bibitem[{{Shakeri} {et~al.}(2020){Shakeri}, {Hajkarim}, \& {Xue}}]{2020JHEP...12..194S}
{Shakeri}, S., {Hajkarim}, F., \& {Xue}, S.-S. 2020, Journal of High Energy Physics, 2020, 194, \dodoi{10.1007/JHEP12(2020)194}

\bibitem[{Shaposhnikov(2008)}]{nuMSM}
Shaposhnikov, M. 2008, JHEP, 08, 008, \dodoi{10.1088/1126-6708/2008/08/008}

\bibitem[{{Smirnov} \& {Trautner}(2023)}]{2023PhRvL.131b1002S}
{Smirnov}, A.~Y., \& {Trautner}, A. 2023, \prl, 131, 021002, \dodoi{10.1103/PhysRevLett.131.021002}

\bibitem[{{Volonteri}(2012)}]{2012Sci...337..544V}
{Volonteri}, M. 2012, Science, 337, 544, \dodoi{10.1126/science.1220843}

\bibitem[{{Volonteri} {et~al.}(2021){Volonteri}, {Habouzit}, \& {Colpi}}]{2021NatRP...3..732V}
{Volonteri}, M., {Habouzit}, M., \& {Colpi}, M. 2021, Nature Reviews Physics, 3, 732, \dodoi{10.1038/s42254-021-00364-9}

\bibitem[{{Woods} {et~al.}(2019){Woods}, {Agarwal}, {Bromm}, \& {et al.}}]{2019PASA...36...27W}
{Woods}, T.~E., {Agarwal}, B., {Bromm}, V., \& {et al.} 2019, \pasa, 36, e027, \dodoi{10.1017/pasa.2019.14}

\bibitem[{{Woods} {et~al.}(2017){Woods}, {Heger}, {Whalen}, {Haemmerl{\'e}}, \& {Klessen}}]{2017ApJ...842L...6W}
{Woods}, T.~E., {Heger}, A., {Whalen}, D.~J., {Haemmerl{\'e}}, L., \& {Klessen}, R.~S. 2017, \apjl, 842, L6, \dodoi{10.3847/2041-8213/aa7412}

\bibitem[{{Yang} {et~al.}(2021){Yang}, {Wang}, {Fan}, \& {et al.}}]{2021ApJ...923..262Y}
{Yang}, J., {Wang}, F., {Fan}, X., \& {et al.} 2021, \apj, 923, 262, \dodoi{10.3847/1538-4357/ac2b32}

\bibitem[{{Yue} {et~al.}(2023){Yue}, {Eilers}, {Simcoe}, {Mackenzie}, {Matthee}, {Kashino}, {Bordoloi}, {Lilly}, \& {Naidu}}]{2023arXiv230904614Y}
{Yue}, M., {Eilers}, A.-C., {Simcoe}, R.~A., {et~al.} 2023, arXiv e-prints, arXiv:2309.04614, \dodoi{10.48550/arXiv.2309.04614}

\bibitem[{{Yunis} \& et~al.(2020)}]{2020PDU....3000699Y}
{Yunis}, R., \& et~al. 2020, Phys. Dark Universe, 30, 100699, \dodoi{https://doi.org/10.1016/j.dark.2020.100699}

\bibitem[{{Zhang} {et~al.}(2022){Zhang}, {Abdukerim}, {Bo}, {Chen}, {Chen}, {Chen}, {Cheng}, {Cheng}, {Cui}, {Fan}, {Fang}, {Fu}, {Fu}, {Geng}, {Giboni}, {Gu}, {Guo}, {Han}, {He}, {He}, {Huang}, {Huang}, {Huang}, {Hou}, {Ji}, {Ju}, {Li}, {Li}, {Li}, {Li}, {Li}, {Lin}, {Liu}, {Lu}, {Luo}, {Luo}, {Ma}, {Ma}, {Mao}, {Shaheed}, {Meng}, {Ning}, {Qi}, {Qian}, {Ren}, {Shang}, {Shang}, {Shen}, {Si}, {Sun}, {Tan}, {Tao}, {Wang}, {Wang}, {Wang}, {Wang}, {Wang}, {Wang}, {Wang}, {Wang}, {Wei}, {Wu}, {Wu}, {Xia}, {Xiao}, {Xiao}, {Xie}, {Yan}, {Yan}, {Yang}, {Yang}, {Yu}, {Yuan}, {Yuan}, {Zeng}, {Zhang}, {Zhang}, {Zhang}, {Zhang}, {Zhang}, {Zhang}, {Zhao}, {Zheng}, {Zhou}, {Zhou}, {Zhou}, {Zhou}, {Zhou}, {Ge}, {He}, {Ma}, {Sheng}, \& {PandaX Collaboration}}]{2022PhRvL.129p1804Z}
{Zhang}, D., {Abdukerim}, A., {Bo}, Z., {et~al.} 2022, \prl, 129, 161804, \dodoi{10.1103/PhysRevLett.129.161804}

\bibitem[{{Zhu} {et~al.}(2022){Zhu}, {Li}, {Li}, {Maji}, {Yajima}, {Schneider}, \& {Hernquist}}]{2022MNRAS.514.5583Z}
{Zhu}, Q., {Li}, Y., {Li}, Y., {et~al.} 2022, \mnras, 514, 5583, \dodoi{10.1093/mnras/stac1556}

\end{thebibliography}

\end{document}